\documentstyle[12pt]{article}
\newcommand{\bold}[1]{\mbox{\boldmath ${#1}$}}
\newcommand{\la}{\langle}
\newcommand{\ra}{\rangle}
\newcommand{\deut}{\langle \mbox{\boldmath $r$} | \Phi^{d}_{(1)M} \rangle}
\newcommand{\ppwave}[1]
{\langle {#1} | \Phi^{2p \ \mbox{\scriptsize {\bf k}}}_{(S)M_{S}} \rangle }
\newcommand{\isodeut}{\langle \mbox{\boldmath $r$} | \Psi^{d}_{(1)M} \rangle}
\newcommand{\isoppwave}[1]
{\langle {#1} | \Psi^{2p \ \mbox{\scriptsize {\bf k}}}_{(S)M_{S}} \rangle}
\newcommand{\ketdeut}{| \Phi^{d}_{(1)M} \rangle}
\newcommand{\brapp}{\langle \Phi^{2p \ \mbox{\scriptsize {\bf k}}}_{(S)M_{S}} |
}
\newcommand{\bradeut}{\langle \Phi^{d}_{(1)M} |}
\newcommand{\ketpp}{| \Phi^{2p \ \mbox{\scriptsize {\bf k}}}_{(S)M_{S}}
\rangle}
\newcommand{\ketisodeut}{ | \Psi^{d}_{(1)M} \rangle}
\newcommand{\braisodeut}{ \langle \Psi^{d}_{(1)M} |}
\newcommand{\braisopp}
{\langle \Psi^{2p \ \mbox{\scriptsize {\bf k}}}_{(S)M_{S}} |}
\newcommand{\ketisopp}
{| \Psi^{2p \ \mbox{\scriptsize {\bf k}}}_{(S)M_{S}} \rangle}
\renewcommand{\theequation}{\arabic{section}.\arabic{equation}}
\begin{document}
\title{ Quasi-Elastic $d(\vec{p},\vec{n})2p$ Reaction and Spin Response
Functions
of the Deuteron }
\author{Atsushi ITABASHI, Kazunori AIZAWA$^{*}$ and Munetake ICHIMURA \\
{\it Institute of Physics, University of Tokyo, Tokyo 153 }\\
$^{*}${\it Hitachi, Ltd., Kawasaki 211} }
\date{ }
\setlength{\oddsidemargin}{0cm}
\setlength{\baselineskip}{7mm}
\maketitle
\begin{abstract}
We calculated response functions of the deuteron for charge exchange
processes, including
the final state interaction between two protons. Using them we evaluated the
double differential cross section and polarization observables of
$d(\vec{p},\vec{n})2p$ by means of plane wave impulse approximation with an
optimal factorization. Calculation
well reproduced the shape of the energy spectra of the cross section, though
somewhat overestimated the magnitude. It also reproduced the spin observables
well.
\end{abstract}
\vspace{5mm}
\begin{center}
{\bf UT-Komaba 93-20}
\end{center}
\vspace{5mm}
\setcounter{equation}{0}
\section{Introduction}
Intermediate energy nucleon-nucleus scattering with large momentum transfer
($|\bold{q}| \ \mbox{\raisebox{-4pt}{$\stackrel{>}{\sim}$}} \ 1$ fm$^{-1}$)
shows
a broad
bump in the energy spectrum.
This is called a quasi-elastic
scattering because the scattering from an
individual particle in the nucleus is expected to play a dominant role in
the process. The peak is observed near the nucleon recoil energy
$\bold{q}^2 / 2M_{N}$($M_{N}$ being the nucleon mass)
and its broad width is thought to reflect the Fermi motion.
\par
For an external field which induces charge exchange processes, the nuclear
response is separated into those for
isovector spin-scalar $ \bold{\tau} $, isovector spin-longitudinal
$ \bold{\tau} (\bold{\sigma} \cdot \hat{\bold{q}}) $, and
isovector spin-transverse
$ \bold{\tau} (\bold{\sigma} \times \hat{\bold{q}}) $
 modes.
We write the response functions for these modes as $R_{S}$, $R_{L}$, and
$R_{T}$
respectively. It is a very
interesting subject of recent nuclear physics to extract them
from the quasi-elastic scatterings.
In the early 80's Alberico et al.\cite{alberico} predicted that the precursor
phenomena of the pion condensation are
prominent in this region. Their calculation showed that in the high
$\bold{q}$ transfer region $R_{L}$ is enhanced and $R_{T}$ is
quenched relative to those
in the free Fermi gas.
Therefore the ratio $R_{L}/R_{T}$ should be much
larger than 1.\par
Since the polarimeter technique has improved
greatly these days, complete polarization
transfer experiments have become feasible. Now the observables like the
depolarization tensor $D_{ij}$ as well as the analyzing power $A_{y}$ are
available. The prediction of Alberico et al. inspired experimentalists to
extract $R_{L}$ and $R_{T}$ from such polarization observables.\par
The transverse response
function $R_{T}$ has been measured in a wide range of $(\omega,\bold{q})$
by electron scatterings \cite{altemus}
for a long time($\omega$ being the energy transfer). However we need to
use the hadron probe instead to
excite the  spin-longitudinal mode. The first experiment which extracted
the response in this mode was performed by Carey et al.\cite{carey} at
Los Alamos meson physics facility(LAMPF).
They carried out the inclusive $(p,p')$
scattering experiment using 500 MeV proton. The targets were $\ ^{2}$H,
Ca, Pb.  However they got the result of $R_{L}/R_{T} \
\mbox{\raisebox{-4pt}{$\stackrel{<}{\sim}$}} \ 1$, which
contradicted the prediction.  \par
The ($p$,$p'$) reaction does not distinguish the isoscalar response from
the isovector one, while the ($p$,$n$) reaction makes a good probe of
$R_{L}$ because of its purely isovector nature. Lately Chen et al.\cite{chen}
performed polarization transfer measurements for the quasi-elastic
($\vec{p}$,$\vec{n}$) reaction at LAMPF with the incident energy 495 MeV at
the laboratory angle
$\theta_{lab}=18^{\circ}$.
The targets were $\ ^{2}$H,
$\ ^{12}$C, $\ ^{40}$Ca.
Further experimental studies are now under way for various angles and
bombarding energies at LAMPF and the Research Center for Nuclear Physics, Osaka
University(RCNP) \cite{sakai}. \par
In the analysis of these experiments the data of the deuteron played the
essential
role, the role of reference. When Carey et al. derived the ratio $R_{L}/R_{T}$,
they assumed $R^{d}_{L}/R^{d}_{T}=1$($d$ referring to the deuteron)
regarding the deuteron as an assembly of a
free proton and neutron. However there is a possibility that the correlations
among
constituent nucleons, such as the well-known tensor correlation, are important
for determining the ratio. Very recently Pandharipande et al.\cite{pandhari}
calculated
the response functions $R^{d}_{L}$ and $R^{d}_{T}$ including such correlations
in the
initial deuteron and final $2p$ states by using Urbana-Argonne interaction.
They
found that the ratio $R^{d}_{L}/R^{d}_{T}$ is close to 1 at the peak region but
deviate rather much from 1 at high and low $\omega$.
In this paper we have calculated them together with $R^{d}_{S}$ making use of
the Reid soft core
potential \cite{reid}. We have obtained the same result as theirs on the
ratio(see sect.4).
\par
We also need to know the reaction mechanism for extracting the response
functions
from the experimental data. We have considered $d(\vec{p},\vec{n})2p$ reaction,
because
the deuteron gives a good test case for that purpose. Since it is a weakly
bound two
nucleon system, we expect simple mechanism for such intermediate energy
reacitons.
Furthermore reliable calculations are feasible in this nuclear system. \par
Early theoretical studies of this reaction were done by
Phillips \cite{phillips}, Dass and Queen \cite{dass}. They calculated $D_{nn}$
at the incident energy of 30 $\sim$ 160 MeV with inclusion of
only S-wave correlation of the final $2p$.
Ramavataram and Ho-Kim \cite{ramavataram} treated the reaction at the
incident energy of 25 $\sim$ 100 MeV with
different sets of {\it NN} phase shift. As the polarization
measurements improved in the intermediate energies, theoretical studies also
made
progress. In this energy region
Bugg and Wilkin investigated the tensor polarization of the $p(\vec{d},2p)n$
reaction where the outgoing two protons has very low
excitation energy \cite{buggwilkin1}.
Carbonell, Barbaro and
Wilkin also considered the same reaction in a more advanced scheme
 \cite{carbonell}. Recently Deloff and Siemiarczuk \cite{deloff}
considered relatively high
energy scattering by the impulse approximation.
They treated $pd \rightarrow pnp$ reaction at 1 GeV including the
$N\Delta$ channel in {\it NN} scattering. \par
Here we have taken the plane wave impulse approximation(PWIA), and
followed the optimal factorization
formalism described by Ichimura and Kawahigashi \cite{ichimura} and the
method of partial wave development of Carbonell, Barbaro and
Wilkin \cite{carbonell}.
We have calculated the
unpolarized double differential cross section and the polarization quantities
of the
reaction $d(\vec{p},\vec{n})2p$ performed by Chen et al.\cite{chen}
with the response functions mentioned above. \par
In sect.2 we present the formalism to calculate the {\it t}-matrix of
the reaction in PWIA with the optimal factorization. In sect.3 we define the
response
functions and relate them to polarization observables. In sect.4 we
give the results and the discussions.
\setcounter{equation}{0}
\section{Formalism}
\subsection{{\it t}-matrix of $d(p,n)2p$}
In the impulse approximation we take the transition operator
of this reaction $\hat{T}$ as the sum of the {\it NN} transition
operator $\hat{t}_{0 i}$. Here we refer to 0 as the incident particle
and $i$ as the $i$-th nucleon in the target.
The deuteron is a weak binding system and its density is
comparatively low. In addition the projectile has kinetic energy of the order
of
several hundred MeV. So one may expect that the PWIA is reasonable for this
target, the confirmation of which is one of our main purposes.
\par
The {\it t}-matrix element of the reaction can
be written in PWIA as
\begin{eqnarray}
     \la f | \hat{T} | i \ra
 &=& \int d \bold{k}_{1} d \bold{k'}_{1}
     \delta^{(3)}( \bold{k}_{f} + \bold{k'}_{1} - \bold{k}_{i} -
     \bold{k}_{1} )
     \int d \bold{k}_{2} d \bold{k'}_{2}
     \delta^{(3)}(\bold{k}_{2} - \bold{k'}_{2}) \nonumber \\
 & & \times \la \chi_{\mu_{f}} ; n |
     \la \varphi_{f} | \bold{k'}_{1} \bold{k'}_{2} \ra
     \la \bold{k}_{f} \bold{k'}_{1} | \hat{t}_{01} |
     \bold{k}_{i} \bold{k}_{1} \ra
     \la \bold{k}_{1} \bold{k}_{2} | \varphi_{i} \ra  | \chi_{\mu_{i}} ; p\ra
     \nonumber \\
 & & \mbox{} +( 1 \longleftrightarrow 2 ) \nonumber\\
 &=& 2 \int d \bold{r}_{1} d \bold{r'}_{1}
     \int d \bold{r}_{2} d \bold{r'}_{2}
     \delta^{(3)}( \bold{r}_{2} - \bold{r'}_{2} )
     \int \frac{d \bold{p}}{(2 \pi)^{3}}
     e^{i \bold{p} ( \bold{r'}_{1} - \bold{r}_{1} )}
      e^{-i \bold{k}_{f} \bold{r'}_{1} } \la \chi_{\mu_{f}} ; n | \nonumber \\
 & & \times
     \la \varphi_{f} | \bold{r'}_{1} \bold{r'}_{2} \ra
     \la \bold{k}_{f} \ \bold{p}-\bold{k}_{f} | \hat{t}_{01} |
     \bold{k}_{i} \ \bold{p}-\bold{k}_{i} \ra
     \la \bold{r}_{1} \bold{r}_{2} | \varphi_{i} \ra
     | \chi_{\mu_{i}} ; p \ra e^{i \bold{k}_{i} \bold{r}_{1}}
      \label{1A}
\end{eqnarray}
where $\varphi_{i}$ and $\varphi_{f}$ are
the initial and final states of the target, $| \chi_{\mu_{i}} ; p \ra $ and
$ | \chi_{\mu_{f}} ; n \ra $
are the spin and isospin states of the incident proton and the outgoing neutron
respectively.
We separate the center-of-mass(c.m.) motion from
$ \la \bold{r}_{1} \bold{r}_{2} | \varphi_{i} \ra$
and $\la \bold{r'}_{1} \bold{r'}_{2} | \varphi_{f} \ra$ as
\begin{eqnarray}
     \la \bold{r}_{1} \bold{r}_{2} | \varphi_{i} \ra
 &=& \la \bold{r} | \Psi^{d}_{(1) M} \ra
          e^{i \bold{K}_{i} \bold{R} }, \quad
     \la \bold{r'}_{1} \bold{r'}_{2} | \varphi_{f} \ra
 =  \isoppwave{\bold{r}'} e^{i \bold{K}_{f} \bold{R'} }. \label{1B}
\end{eqnarray}
Here $ \bold{R} $ and $ \bold{r} $ are the c.m.\
and the relative coordinates;
likewise for $ \bold{R}' $ and $ \bold{r}' $. The subscripts of $\Psi$ mean the
total
intrinsic spin and its projection.
The final $2p$ does not have a bound state and is specified by the aymptotic
relative momentum $\bold{k}$ and its spin ($S,M_{S}$).
Performing some integrations we obtain
\begin{equation}
 \la f | \hat{T} | i \ra = (2 \pi)^{3}
    \delta^{(3)}(\bold{k}_{i} + \bold{K}_{i} - \bold{k}_{f} - \bold{K}_{f})
    \tilde{T}_{f i} ,   \label{1C}
\end{equation}
\begin{eqnarray}
 \tilde{T}_{fi}
 &=& 2
     \int d \bold{r} d \bold{r'}
     \int \frac{d \bold{p}}{(2 \pi)^{3}}
     \la \chi_{\mu_{f}} ; n | \braisopp\bold{r'} \ra
     \la \bold{k}_{f} \ \bold{p}-\bold{k}_{f} | \hat{t}_{01} |
     \bold{k}_{i} \ \bold{p}-\bold{k}_{i} \ra
     \la \bold{r} | \Psi^{d}_{(1) M} \ra  | \chi_{\mu_{i}} ; p \ra \nonumber \\
 & & \times
     e^{i (  \frac{1}{2}\bold{K}_{f} + \bold{k}_{a} - \bold{p} )
     \bold{r} }
     e^{-i ( \frac{1}{2} \bold{K}_{f} + \bold{k}_{a} +
     \frac{1}{2}\bold{q} - \bold{p} )\bold{r}' } \label{1D}
\end{eqnarray}
with the momentum transfer $\bold{q}$ and the average
momentum $\bold{k}_{a}$,
\begin{eqnarray}
\bold{q}     &=& \bold{k}_{f} - \bold{k}_{i}, \quad
\bold{k}_{a} = \frac{\bold{k}_{i} + \bold{k}_{f}}{2}. \label{1E}
\end{eqnarray}
It is very complicated to advance the calculation, because there
is $\bold{p}$ dependence in the {\it NN t}-matrix, and it is
off-energy-shell in general. To proceed furthur we replace $ \bold{p} $
in the {\it NN t}-matrix with a certain fixed value $ \tilde{\bold{p}} $
which well represent the contribution to the integral.
Equation (\ref{1C}) becomes after the integration over $\bold{r}'$ and
$\bold{p}$
\begin{eqnarray}
 \tilde{T}_{fi}
 = 2 \int d \bold{r}
     \la \chi_{\mu_{f}} ; n | \braisopp \bold{r} \ra
     \la \bold{k}_{f} \ \tilde{\bold{p}}-\bold{k}_{f} | \hat{t}_{01} |
     \bold{k}_{i} \ \tilde{\bold{p}}-\bold{k}_{i} \ra
     \la \bold{r} | \Psi^{d}_{(1) M} \ra  | \chi_{\mu_{i}} ; p \ra
     e^{ - \frac{i}{2} \bold{q} \bold{r}} .  \label{1G}
\end{eqnarray}
\par
{}From now on we will evaluate
$\tilde{T}_{fi}$ in the projectile-target c.m.\ frame.
Following the prescription of optimal factorization \cite{ichimura} \cite{zhu}
we set the optimum value $ \tilde{\bold{p}} $ as
\begin{equation}
 \tilde{\bold{p}} = \frac{\bold{k}_{a}}{2} - \eta \bold{q}, \label{1H}
\end{equation}
where $\eta$ is
determined by the on-energy-shell condition
\begin{equation}
  E(\bold{k}_{f}) + E(\tilde{\bold{p}} - \bold{k}_{f})
  = E(\bold{k}_{i}) + E(\tilde{\bold{p}} - \bold{k}_{i}) \label{1K}
\end{equation}
with $E(\bold{k}) = \sqrt{M_{N}^2 + \bold{k}^2}$.
The {\it NN t}-matrix in Eq.(\ref{1G}) now becomes
\begin{equation}
 \la \bold{k}_{a} + \frac{1}{2} \bold{q},(- \frac{1}{2} - \eta)\bold{q}
   - \frac{1}{2} \bold{k}_{a}
   | \hat{t}_{01} |
   \bold{k}_{a} - \frac{1}{2} \bold{q},(\frac{1}{2} - \eta)\bold{q}
   - \frac{1}{2} \bold{k}_{a} \ra \label{1J}
\end{equation}
and we call this the {\it NN t}-matrix in the $\eta$-frame. \\
\subsection{ Observables }
We introduce the orthogonal unit vectors in the $p$-$d$ c.m. system as
\begin{equation}
 \hat{\bold{q}} = \frac{\bold{q}}{|\bold{q}|}, \quad
 \hat{\bold{n}} = \frac{\bold{k}_{i} \times \bold{k}_{f}}
                  {|\bold{k}_{i} \times \bold{k}_{f}|}, \quad
 \hat{\bold{p}} = \hat{\bold{q}} \times \hat{\bold{n}}. \label{1M}
\end{equation}
The unpolarized differential cross
section $I_{0}$ and the depolarization tensor $D_{ij}$
are expressed as
\begin{eqnarray}
 I_{0} &=& K \mbox{Tr} [ \tilde{T} \tilde{T}^{\dagger} ], \quad
 D_{ij} = \frac{ \mbox{Tr} [\tilde{T} \sigma_{0i} \tilde{T}^{\dagger}
  \sigma_{0j}]}
 { \mbox{Tr} [ \tilde{T} \tilde{T}^{\dagger} ]}, \label{1N}
\end{eqnarray}
where the subscripts {\it i} and {\it j}
represent one of the directions $\hat{\bold{q}}$, $\hat{\bold{n}}$
 and $\hat{\bold{p}}$, whereas the subscript 0 indicates the projectile.
In the equations above
$\mbox{Tr} [\hat{\cal O}_{1} \hat{\cal O}^{\dagger}_{2}]$ means
\begin{eqnarray}
\mbox{Tr} [\hat{\cal O}_{1} \hat{\cal O}^{\dagger}_{2}]
&=& \sum_{\mu_{i} \mu_{f}}\sum_{M} \sum_{S M_{S}} \int d\bold{k}
   \la \chi_{\mu_{f}} ; n | \braisopp \hat{\cal O}_{1}
  \ketisodeut |\chi_{\mu_{i}} ; p \ra \nonumber \\
& &  \times \la \chi_{\mu_{i}} ; p | \braisodeut \hat{\cal O}^{\dagger}_{2}
   \ketisopp |\chi_{\mu_{f}} ; n \ra
  \delta( \omega - ( {\cal E}^{f} - {\cal E}^{i} ) ), \nonumber
\end{eqnarray}
where
$\omega = E(\bold{k}_{i}) - E(\bold{k}_{f})$, and ${\cal E}^{i}$
and ${\cal E}^{f}$
are the target energy of the initial and the final states respectively. The
energy of
the deuteron ${\cal E}^{i}$ is given by $\sqrt{ M^{2}_{d} + \bold{k}^{2}_{i}
}$.
We approximate ${\cal E}^{f}$ as
$\sqrt{ (2 M_{p} + \bold{k}^2/M_{p})^{2} + \bold{k}^{2}_{f} }$, because we
consider
only the region where the relative motion of $2p$ can be treated
non-relativistically.
The kinematical factor {\it K} in Eq.(\ref{1N}) is given by
\begin{equation}
 K = \frac{m_{i} m_{f}}{(2 \pi)^{2}} \frac{k_{f}}{k_{i}} \frac{1}{6}
     \label{1Q}
\end{equation}
where $m_{i}$ and $m_{f}$ are the relativistic reduced masses
\begin{eqnarray}
 m_{i} &=& \frac{E(\bold{k}_{i}) {\cal E}^{i}}
                {E(\bold{k}_{i}) + {\cal E}^{i}}, \quad
 m_{f} = \frac{E(\bold{k}_{f}) {\cal E}^{f}}
                {E(\bold{k}_{f}) + {\cal E}^{f}}.  \label{1R}
\end{eqnarray}
\par
The unpolarized double differential cross section in the laboratory frame
$I_{lab}$
is connected to $I_{0}$ through the relation \cite{hagedorn}
\begin{equation}
I_{lab} \sin \theta_{lab} = I_{0} \sin \theta_{c.m.}.
\end{equation}
Bleszynski et al.\cite{blesz} introduced the quantities $D_{i}$ expressed by
$D_{ij}$ as
\begin{eqnarray}
 I_{0} D_{0} &=& \frac{I_{0}}{4} [ 1 + D_{nn} + D_{qq} + D_{pp} ] ,\nonumber \\
 I_{0} D_{n} &=& \frac{I_{0}}{4} [ 1 + D_{nn} - D_{qq} - D_{pp} ] ,\nonumber \\
 I_{0} D_{q} &=& \frac{I_{0}}{4} [ 1 - D_{nn} + D_{qq} - D_{pp} ] ,\nonumber \\
 I_{0} D_{p} &=& \frac{I_{0}}{4} [ 1 - D_{nn} - D_{qq} + D_{pp} ] .\label{1S}
\end{eqnarray}
We will show the results in terms of them. \\
\subsection{ {\it NN t}-matrix }
In Eq.(\ref{1G}) $\tilde{T}_{f i}$
includes the {\it NN t}-matrix in the $\eta$-frame, which we should evaluate
from
the one in the {\it NN} c.m.\ frame
$\la \bold{\kappa}_{f}, - \bold{\kappa}_{f}| \hat{t}_{01} |
   \bold{\kappa}_{i}, - \bold{\kappa}_{i} \ra$.
This is described in the convention of Kerman, McManus, and Thaler \cite{KMT}
as
\begin{eqnarray}
 \la \bold{\kappa}_{f}, - \bold{\kappa}_{f}| \hat{t}_{01} |
 \bold{\kappa}_{i}, - \bold{\kappa}_{i} \ra
 &=& - \frac{4 \pi}{E(\bold{\kappa})} M(\bold{\kappa}_{i},\bold{\kappa}_{f})
 \nonumber \\
 M(\bold{\kappa}_{i},\bold{\kappa}_{f})
 &=& A + B(\bold{\sigma^{(1)}} \cdot \hat{\bold{n}}_{c})
          (\bold{\sigma^{(0)}} \cdot \hat{\bold{n}}_{c})
       + C(\bold{\sigma^{(1)}} \cdot \hat{\bold{n}}_{c} +
           \bold{\sigma^{(0)}} \cdot \hat{\bold{n}}_{c}) \nonumber \\
 & &   + E(\bold{\sigma^{(1)}} \cdot \hat{\bold{q}}_{c})
          (\bold{\sigma^{(0)}} \cdot \hat{\bold{q}}_{c})
       + F(\bold{\sigma^{(1)}} \cdot \hat{\bold{p}}_{c})
          (\bold{\sigma^{(0)}} \cdot \hat{\bold{p}}_{c}) \label{2C}
\end{eqnarray}
with the orthogonal unit vectors
\begin{equation}
 \hat{\bold{q}}_{c} = \frac{\bold{\kappa}_{f} - \bold{\kappa}_{i}}
                      {|\bold{\kappa}_{f} - \bold{\kappa}_{i}|}, \quad
 \hat{\bold{n}}_{c} = \frac{\bold{\kappa}_{i} \times \bold{\kappa}_{f}}
                      {|\bold{\kappa}_{i} \times \bold{\kappa}_{f}|}, \quad
 \hat{\bold{p}}_{c} = \hat{\bold{q}}_{c} \times \hat{\bold{n}}_{c} .
   \label{2D}
\end{equation}
Here we used $ |\bold{\kappa}_{i}| = |\bold{\kappa}_{f}| = |\bold{\kappa}| $
and
$ M(\bold{\kappa}_{i},\bold{\kappa}_{f}) $ is the {\it NN} scattering
amplitude. The coefficients $A \sim F$ are determined by the {\it NN}
scattering
experiments. Due to charge
independence they are written with the isospin operator as
\begin{equation}
 A = A_{0} + A_{1} (\bold{\tau}^{(0)} \cdot \bold{\tau}^{(1)}),\ \mbox{etc.}
\label{2E}
\end{equation}
\par
The relation between the {\it NN t}-matrices in the $\eta$-frame and in the
{\it NN} c.m.\ frame is
\begin{eqnarray}
& & \la \bold{k}_{a} + \frac{1}{2} \bold{q},(- \frac{1}{2}-\eta)\bold{q}
- \frac{1}{2} \bold{k}_{a}
   | \hat{t}_{01} |
   \bold{k}_{a} - \frac{1}{2} \bold{q},( \frac{1}{2}-\eta)\bold{q}
- \frac{1}{2} \bold{k}_{a} \ra  \label{2F} \\
   & = & J(\bold{k}_{a},\bold{q},\eta)
       \exp(\frac{i}{2}\chi' \sigma_{0n})
       \exp(-\frac{i}{2}\rho' \sigma_{1n})
       \la \bold{\kappa}_{f}, - \bold{\kappa}_{f}| \hat{t}_{01} |
       \bold{\kappa}_{i}, - \bold{\kappa}_{i} \ra
       \exp( \frac{i}{2} \chi' \sigma_{0n})
       \exp(-\frac{i}{2}\rho \sigma_{1n})  \nonumber
\end{eqnarray}
Here $J(\bold{k}_{a},\bold{q},\eta)$ is a M\"oller factor
\begin{equation}
 J(\bold{k}_{a},\bold{q},\eta) = \frac{E(\bold{\kappa})^2}
                                 {\sqrt{
                                 E( \bold{k}_{a} + \frac{1}{2} \bold{q} )
                 E( (-\eta-\frac{1}{2})\bold{q}-\frac{1}{2} \bold{k}_{a} )
                                 E( \bold{k}_{a} - \frac{1}{2} \bold{q} )
                 E( (-\eta+\frac{1}{2})\bold{q}-\frac{1}{2} \bold{k}_{a} )}}.
  \label{2G}
\end{equation}
and $\chi$, $\chi'$, $\rho$, and $\rho'$ are spin rotation angles
defined in Ref.\ \cite{ichimura}. \par
Now we introduce two approximations.
One is to identify the unit vector $\hat{\bold{q}}$ in the
$\eta$-frame with $\hat{\bold{q}_{c}}$ because
the angle $\psi$ between them is extremely small for wide range of $\omega$ as
is
shown in Table I. Since we determine the optimal value
by Eq.(\ref{1H}), the reaction
plane is common for the $\eta$ frame and the {\it NN} c.m.\ frame. Then
\begin{equation}
 \hat{\bold{q}} \approx \hat{\bold{q}}_{c}, \quad
 \hat{\bold{n}} \approx \hat{\bold{n}}_{c}, \quad
 \hat{\bold{p}} \approx \hat{\bold{p}}_{c} .
 \label{2H}
\end{equation}
The other approximation is to neglect the spin rotation, because the rotation
angles
are very small in the present case as shown
in Table I. The {\it NN t}-matrix element in the $\eta$-frame now turns to be
\begin{eqnarray}
& & \la \bold{k}_{a} + \frac{1}{2} \bold{q},(- \frac{1}{2}-\eta)\bold{q}
- \frac{1}{2} \bold{k}_{a}
   | \hat{t}_{01} |
   \bold{k}_{a} - \frac{1}{2} \bold{q},( \frac{1}{2}-\eta)\bold{q}
- \frac{1}{2} \bold{k}_{a} \ra \nonumber \\
  & \approx &  J(\bold{k}_{a},\bold{q},\eta)
       \la \bold{\kappa}_{f}, - \bold{\kappa}_{f}| t_{01} |
       \bold{\kappa}_{i}, - \bold{\kappa}_{i} \ra. \label{2I}
\end{eqnarray}
Then we express the {\it t}-matrix (\ref{1G}) as
\begin{eqnarray}
  \tilde{T}_{f i}
 &=& 2 \zeta
     \left[  A_{1} \la \chi_{\mu_{f}};n|\bold{\tau}^{(0)} |
     \chi_{\mu_{i}};p \ra
     \braisopp \bold{\tau}^{(1)}
         e^{- \frac{i}{2} \bold{q} \cdot \bold{r}}
     \ketisodeut \right. \nonumber \\
 & & +  B_{1} \la \chi_{\mu_{f}};n | \bold{\tau}^{(0)}
     (\bold{\sigma}^{(0)} \cdot \hat{\bold{n}}) |\chi_{\mu_{i}};p \ra
     \braisopp \bold{\tau}^{(1)}
     (\bold{\sigma}^{(1)} \cdot \hat{\bold{n}})
     e^{- \frac{i}{2} \bold{q} \cdot \bold{r}}
    \ketisodeut  \nonumber \\
 & & +  C_{1} \la \chi_{\mu_{f}};n| \bold{\tau}^{(0)}
     (\bold{\sigma}^{(0)} \cdot \hat{\bold{n}})|\chi_{\mu_{i}};p \ra
     \braisopp \bold{\tau}^{(1)}
     e^{- \frac{i}{2} \bold{q} \cdot \bold{r}}
     \ketisodeut \nonumber \\
 & & +  C_{1} \la \chi_{\mu_{f}};n| \bold{\tau}^{(0)} |\chi_{\mu_{i}};p \ra
     \braisopp \bold{\tau}^{(1)}
     (\bold{\sigma}^{(1)} \cdot \hat{\bold{n}})
     e^{- \frac{i}{2} \bold{q} \cdot \bold{r}}
     \ketisodeut  \label{2J} \\
 & & +  E_{1} \la \chi_{\mu_{f}};n| \bold{\tau}^{(0)}
     (\bold{\sigma}^{(0)} \cdot \hat{\bold{q}})|\chi_{\mu_{i}};p \ra
     \braisopp \bold{\tau}^{(1)}
     (\bold{\sigma}^{(1)} \cdot \hat{\bold{q}})
     e^{- \frac{i}{2} \bold{q} \bold{r}}
     \ketisodeut  \nonumber \\
 & & + \left.  F_{1} \la \chi_{\mu_{f}};n| \bold{\tau}^{(0)}
     (\bold{\sigma}^{(0)} \cdot \hat{\bold{p}})|\chi_{\mu_{i}};p \ra
     \braisopp \bold{\tau}^{(1)}
     (\bold{\sigma}^{(1)} \cdot \hat{\bold{p}})
     e^{- \frac{i}{2} \bold{q} \bold{r}}
     \ketisodeut
     \right]   \nonumber
\end{eqnarray}
with
\begin{equation}
\zeta = -J(\bold{k}_{a}, \bold{q}, \eta) \frac{4 \pi}{E(\bold{\kappa})}
\label{2K}
\end{equation}
where we used Eqs.(\ref{2C}), (\ref{2E}), (\ref{2H}) and (\ref{2I}).
\\
\subsection{ Wave function of the target nucleus }
We describe the wave function of the deuteron as
\begin{eqnarray}
 \isodeut & = & \deut
 \frac{ | n \  p \ra - | p \ n \ra }{\sqrt{2}},  \label{3A} \\
 \deut & = & \sum_{\lambda = 0,2 } R_{\lambda}(r)
   [Y_{\lambda}(\Omega_{r}) \times \chi_{1} ]^{(1)}_{M} \label{3B}
\end{eqnarray}
where $Y_{l m}(\Omega_{r})$ is a spherical harmonics and $\chi_{S M_{S}}$
is a spin wave function of the two nucleon system with the spin $S$ and its
projection $M_{S}$.
\par
The scattering wave $\isoppwave{\bold{r}}$
is expressed as
\begin{eqnarray}
 \isoppwave{\bold{r}}
&=&\ppwave{\bold{r}} | p \  p \ra,  \label{3E} \\
 \ppwave{\bold{r}}
&=& \frac{2}{\sqrt{\pi}}
 \sum_{ J M_{J}} \sum_{l m} \sum_{l'+S=\mbox{\scriptsize even}} i^{l} e^{i
\sigma_{l}}
 \frac{ \psi^{JS}_{l',l}(r) }{ kr}
 (l m S M_{S} | J M_{J}) \ Y^{*}_{l m}(\Omega_{k}) \nonumber \\
& & \times [Y_{l'}(\Omega_{r}) \times \chi_{S'}]^{(J)}_{M_{J}},  \label{3F}
\end{eqnarray}
where $\sigma_{l}$ is a Coulomb phase shift. The sum over $l'$ is taken for
$l'+S=$even because of the identical protons. We solve the Schr\"odinger
equation
for $\psi^{JS}_{l',l}(r)$ with the asymptotic form
\begin{eqnarray}
 \psi^{JS}_{l' ,l }(r) & \longrightarrow &
 \frac{ i }{ 2 }
 \left\{
 \delta_{l l'} \exp(-i( k r - \frac{l' \pi}{2} - \eta \ln 2kr + \sigma_{l'}))
 \right. \nonumber \\
 & & \left. - {\cal S}^{JS}_{l' ,l }
 \exp( i( k r - \frac{l' \pi}{2} - \eta \ln 2kr + \sigma_{l'}))
 \right\}
 \label{3I}
\end{eqnarray}
where $\eta$ is a Sommerfeld parameter.
Due to time reversal invariance the S-matrix ${\cal S}$ is
symmetric unitary matrix which can be diagonalized as
\begin{equation}
 {\cal S}^{JS}_{l',l } = \sum_{\alpha} U^{JS}_{\alpha,l' }
 e^{2 i \delta^{J}_{\alpha}} U^{JS}_{\alpha,l }  \label{3J}
\end{equation}
with a real orthogonal matrix $U$.
Using the wave function $\psi^{JS}_{l' ,\alpha}(r)
\equiv \sum_{l'' } U^{JS}_{\alpha,l'' } \psi^{JS}_{l' ,l'' }(r)$, we rewrite
$\ppwave{\bold{r}}$ as
\begin{eqnarray}
 \ppwave{\bold{r}} & = &
 \frac{2}{\sqrt{\pi}}
 \sum_{J M_{J}} \sum_{\alpha} \sum_{l m} \sum_{l'+S = \mbox{\scriptsize even}}
  i^{l} e^{ i \sigma_{l} } U^{JS}_{\alpha,l }
 \frac{ \psi^{JS}_{l' ,\alpha}(r) }{ kr } \nonumber \\
 & & \times (l m S M_{S} | J M_{J}) Y^{*}_{l m}(\Omega_{k})
 [Y_{l'}(\Omega_{r}) \times \chi_{S}]^{(J)}_{M_{J}} . \label{3M}
\end{eqnarray}
\setcounter{equation}{0}
\section{Spin Response Functions}
In the calculation of the observables (\ref{1N}) by Eq.(\ref{2J}) we encounter
the
following four response functions defined as
\begin{eqnarray}
 R_{S}(\bold{q},\omega) &=& \frac{1}{3} \sum_{M} \sum_{S M_{S}}
                           \int d \bold{k}
                           | \braisopp
                           \bold{\tau}^{(1)}
                           e^{ - \frac{i}{2} \bold{q} \cdot \bold{r}}
                           \ketisodeut |^{2} \nonumber \\
                         & & \times
                           \delta(\omega-({\cal E}^{f}-{\cal E}^{i})),
                           \label{4A} \\
 R_{L}(\bold{q},\omega) &=& \frac{1}{3} \sum_{M} \sum_{S M_{S}}
                           \int d \bold{k}
                           | \braisopp
                           \bold{\tau}^{(1)}
                           \bold{\sigma}^{(1)} \cdot \hat{\bold{q}}
                           e^{ - \frac{i}{2} \bold{q} \cdot \bold{r}}
                           \ketisodeut |^{2} \nonumber \\
                        & & \times
                            \delta(\omega-({\cal E}^{f}-{\cal E}^{i})),
                            \label{4B} \\
 R_{T,n}(\bold{q},\omega) &=& \frac{1}{3} \sum_{M} \sum_{S M_{S}}
                           \int d \bold{k}
                           | \braisopp
                           \bold{\tau}^{(1)}
                           \bold{\sigma}^{(1)} \cdot \hat{\bold{n}}
                           e^{ - \frac{i}{2} \bold{q} \cdot \bold{r}}
                           \ketisodeut |^{2} \nonumber \\
                        & & \times
                            \delta(\omega-({\cal E}^{f}-{\cal E}^{i})),
                            \label{4C} \\
 R_{T,p}(\bold{q},\omega) &=& \frac{1}{3} \sum_{M} \sum_{S M_{S}}
                           \int d \bold{k}
                           | \braisopp
                           \bold{\tau}^{(1)}
                           \bold{\sigma}^{(1)} \cdot \hat{\bold{p}}
                           e^{ - \frac{i}{2} \bold{q} \cdot \bold{r}}
                           \ketisodeut |^{2} \nonumber \\
                        & & \times
                            \delta(\omega-({\cal E}^{f}-{\cal E}^{i})).
                            \label{4D}
\end{eqnarray}
They represent the excitation strength of the nucleus
induced by the external field with 4-momentum transfer
($\omega$,$\bold{q}$). Here $R_{S}$ is referred to as the spin-scalar,
$R_{L}$ as the spin-longitudinal, and $R_{T}$ as the spin-transverse response
function. We will see that $R_{T,n} = R_{T,p}$ later. The factor $ \frac{1}{3}
$
means the  average over the initial spin states of the deuteron.
{}From Eqs.(\ref{3A}) and (\ref{3E}) we can easily calculate the isospin part,
for
example,
\begin{eqnarray}
  | \braisopp \bold{\tau}^{(1)}
   (\bold{\sigma}^{(1)} \cdot \hat{\bold{q}})
   e^{ - \frac{i}{2} \bold{q} \bold{r}} \ketisodeut |^{2}
  &=&  | \brapp \bold{\sigma}^{(1)} \cdot \hat{\bold{q}}
             e^{ - \frac{i}{2} \bold{q} \bold{r}} \ketdeut |^{2} . \nonumber
\end{eqnarray}
{}From now on we use the spherical tensor representation such as
\begin{equation}
\sigma_{1 \pm 1}  =  \mp
(\sigma_{x} \pm i \sigma_{y})/\sqrt{2}, \quad
\sigma_{1 0}  =  \sigma_{z}, \quad
\sigma_{0 0}  =  1,
\label{4E}
\end{equation}
\begin{equation}
 q_{1 \pm 1}  =  \mp
(q_{x} \pm i q_{y})/\sqrt{2}, \quad
 q_{1 0}  =  q_{z},\ \mbox{etc.} \label{4F}
\end{equation}
and
\begin{equation}
 \bold{\sigma}^{(1)} \cdot \hat{\bold{q}}
    = \sum_{s_{z}} \sigma^{(1)}_{1 s_{z}} \hat{q}^{\dagger}_{s_{z}}, \
\mbox{etc.}
      \label{4G}
\end{equation}
We can rewrite Eqs.(\ref{4A}) $\sim$ (\ref{4D}) as
\begin{eqnarray}
 R_{S}(\bold{q},\omega) &=& R^{00}_{00}(\bold{q},\omega) , \label{4H} \\
 R_{L}(\bold{q},\omega) &=& \sum_{s_{z} s_{z}'} \hat{q}^{\dagger}_{s_{z}}
                             \hat{q}_{s_{z}'} R^{1 s_{z}}_{1 s_{z}'} ,
                             \label{4I} \\
 R_{T,n}(\bold{q},\omega) &=& \sum_{s_{z} s_{z}'} \hat{n}^{\dagger}_{s_{z}}
                             \hat{n}_{s_{z}'} R^{1 s_{z}}_{1 s_{z}'},
                             \label{4J} \\
 R_{T,p}(\bold{q},\omega) &=& \sum_{s_{z} s_{z}'} \hat{p}^{\dagger}_{s_{z}}
                             \hat{p}_{s_{z}'} R^{1 s_{z}}_{1 s_{z}'} ,
                             \label{4K}
\end{eqnarray}
where
\begin{eqnarray}
 R^{s s_{z}}_{s' s_{z}'}
     & \equiv & \frac{1}{3} \sum_{M} \sum_{S M_{S}}
                \int d \bold{k}
                \bradeut
                \sigma^{(1) \dagger}_{s' s_{z}'}
                e^{ \frac{i}{2} \bold{q} \bold{r}}
                \ketpp
                \brapp
                \sigma^{(1)}_{s s_{z}}
                e^{ - \frac{i}{2} \bold{q} \bold{r}}
                \ketdeut
                \nonumber \\
     & &  \times  \delta(\omega - ( {\cal E}^{f} - {\cal E}^{i} )) .
          \label{4L}
\end{eqnarray}
\par
Using the expressions (\ref{3B}) and (\ref{3M}) we get the matrix
element above as
\begin{eqnarray}
 & & \brapp \sigma^{(1)}_{s s_{z}}
     e^{ - \frac{i}{2} \bold{q} \bold{r} }
    \ketdeut \nonumber \\
 &=& 12 \sum_{\lambda} \sum_{J M_{J}} \sum_{l m} \sum_{L M_{L}}
       \sum_{I M_{I}} \sum_{l'+S = \mbox{\scriptsize even}} \sum_{\alpha}
i^{-L-l}
       (-)^{S-s} \nonumber \\
 & & \times  (l m S M_{S} | J M_{J} ) (L M_{L} s s_{z} | I M_{I})
       (1 M I M_{I} | J M_{J}) (\lambda 0 L 0 | l' 0) \nonumber \\
 & & \times \sqrt{(2I+1)(2L+1)(2 \lambda +1)(2S+1)}
     e^{-i \sigma_{l}} U^{JS}_{\alpha,l } \Gamma^{\alpha l' S J}_{L
\lambda}(k,q)
      Y_{l m}(\Omega_{k}) Y^{\ast}_{L M_{L}}(\Omega_{q}) \nonumber \\
  & & \times \la \frac{1}{2} || \sigma^{(1)}_{s} || \frac{1}{2} \ra
     W( \frac{1}{2} S \frac{1}{2} 1 ; \frac{1}{2} s )
     \left\{ \begin{array}{ccc}
             l'      & S & J \\
             \lambda & 1  & 1 \\
             L       & s  & I
             \end{array}
             \right\} \label{4M}
\end{eqnarray}
where
\begin{equation}
 \la \frac{1}{2} || \sigma^{(1)}_{s} || \frac{1}{2} \ra
 = \left\{ \begin{array}{rl}
           \sqrt{2} & \quad \mbox{for $s=0$} \\
           \sqrt{6} & \quad \mbox{for $s=1$}
           \end{array} \right. \label{4O}
\end{equation}
and $\Gamma^{\alpha l' S J}_{L \lambda}(k,q)$ denotes the integration of
the radial part,
\begin{equation}
 \Gamma^{\alpha l' S J}_{L \lambda}(k,q) = \int r^2 dr
 \frac{ \psi^{JS \ast}_{l',\alpha}(r) }{ kr }  j_{L}(\frac{1}{2} q r)
R_{\lambda}(r).
 \label{4N}
\end{equation}
\par
Now we apply Eq.(\ref{4M}) to Eq.(\ref{4L}). The integration over the
angular part of the relative momentum $\bold{k}$ can be analytically
performed because the argument of the delta function has no angular
dependence due to the approximation for ${\cal E}_{f}$.
By use of the orthogonality of Clebsch-Gordan coefficients,
$U^{JS}_{\alpha,l}$,
and spherical harmonics, $ R^{s s_{z}}_{s' s_{z}'} $ results in
\begin{eqnarray}
 R^{s s_{z}}_{s' s_{z}'} &=& 48 \sum_{L_{1} L_{2}} \sum_{I}
                             \sqrt{ (2L_{1} + 1)(2L_{2} + 1) }
                             \ T^{s s_{z} I}_{s' s_{z}' L_{1} L_{2}}(\Omega_q)
                             \nonumber \\
                         & & \times \int k^2 dk
                             \ \delta(\omega-({\cal E}^{f}-{\cal E}^{i}))
                             \ R^{I}_{s L_{1}, s' L_{2}}(k,q) \label{4P}
\end{eqnarray}
where
\begin{eqnarray}
 T^{s s_{z} I}_{s' s_{z}' L_{1} L_{2}}(\Omega_q)
    &=& \sum_{M_{L_{1}} M_{L_{2}}} \sum_{M_{I}}
        ( L_{1} M_{L_{1}} s s_{z} | I M_{I} ) ( L_{2} M_{L_{2}} s' s_{z}' | I
M_{I} )
        \nonumber \\
    & & \times Y_{L_{1} M_{L_{1}}}^{\ast}(\Omega_{q})
         Y_{L_{2} M_{L_{2}}}(\Omega_{q}),  \label{4Q} \\
 R^{I}_{s L_{1}, s' L_{2}}(k,q)
    &=&    \sum_{S} \sum_{\lambda_{1} \lambda_{2}} \sum_{J}
               \sum_{ l_{1}'+S= \mbox{\scriptsize even}}
               \sum_{ l_{2}'+S= \mbox{\scriptsize even}} \sum_{\alpha}
                i^{L_{2}-L_{1}} (-)^{  s + s' } \nonumber \\
    & & \times  (2J+1)(2S+1) \sqrt{(2 \lambda_{1} +1)(2 \lambda_{2} +1)}
                \Gamma^{\alpha l_{2}' S J \ast }_{L_{2} \lambda_{2} }(k,q)
                \Gamma^{\alpha l_{1}' S J}_{L_{1} \lambda_{1}}(k,q) \nonumber
\\
     & & \times (\lambda_{1} 0 L_{1} 0 | l_{1}' 0)
                (\lambda_{2} 0 L_{2} 0 | l_{2}' 0)
               \la \frac{1}{2} || \sigma^{(1)}_{s} || \frac{1}{2} \ra
               \la \frac{1}{2} || \sigma^{(1)}_{s'} || \frac{1}{2} \ra
\nonumber \\
      & & \times W( \frac{1}{2} S \frac{1}{2} 1 ; \frac{1}{2} s )
               W( \frac{1}{2} S \frac{1}{2} 1 ; \frac{1}{2} s' )
             \left\{ \begin{array}{ccc}
             l_{1}'      & S  & J \\
             \lambda_{1} & 1  & 1 \\
             L_{1}       & s  & I
             \end{array}
             \right\}
            \left\{ \begin{array}{ccc}
             l_{2}'      & S   & J \\
             \lambda_{2} & 1   & 1 \\
             L_{2}       & s'  & I
             \end{array}
             \right\} . \label{4R}
\end{eqnarray}
In this form we are able to separate out $\Omega_{q}$-dependent part. \par
When $ s = s' = 0 $, we obtain
\begin{eqnarray}
 R^{00}_{00} &=& \frac{12}{\pi} \sum_{I} (2I + 1)^{2} \int k^2 dk
                           \delta(\omega-({\cal E}^{f}-{\cal E}^{i})) R^{I}_{0
I,0 I}(k,q),
                           \label{4S} \\
 R^{I}_{0 I,0 I}(k,q) &=& \frac{1}{ 3(2I+1) } \sum_{\lambda_{1} \lambda_{2}}
            \sum_{J} \sum_{ l_{1}'=\mbox{\scriptsize odd}}
            \sum_{ l_{2}'=\mbox{\scriptsize odd}} \sum_{\alpha}
            (2J+1) \sqrt{(2 \lambda_{1} +1)(2 \lambda_{2} +1)} \nonumber \\
    & & \times \Gamma^{\alpha l_{1}' 1 J}_{I \lambda_{1}}(k,q)
            \Gamma^{\alpha l_{2}' 1 J \ast }_{I \lambda_{2} }(k,q)
             (\lambda_{1} 0 I 0 | l_{1}' 0)
             (\lambda_{2} 0 I 0 | l_{2}' 0) \nonumber \\
    & & \times  W( 1 l_{1}' 1 I ; J \lambda_{1} )
            W( 1 l_{2}' 1 I ; J \lambda_{2} ) \label{4T}.
 \end{eqnarray}
\par
In the case of $ s = s' = 1 $ the factor $ R^{I}_{1 L_{1}, 1 L_{2}}(k,q)$ is
rather complicated, so we leave it as it is. On the other hand we can sum
up over $s_{z}$ and $s_{z}'$ in Eqs.(\ref{4I}), (\ref{4J}), and (\ref{4K}).
For the longitudinal mode
\begin{eqnarray}
 \sum_{s_{z} s_{z}'} \hat{q}^{\dagger}_{s_{z}} \hat{q}_{s_{z}'}
 T^{1 s_{z} I}_{1 s_{z}' L_{1} L_{2}}(\Omega_{q})
 &=& \frac{\sqrt{(2L_{1}+1)(2L_{2}+1)}}{4 \pi}
         ( L_{1} 0 1 0 | I 0 ) ( L_{2} 0 1 0 | I 0 )
     \label{4U}
\end{eqnarray}
and for the two transverse modes
\begin{eqnarray}
 \sum_{s_{z} s_{z}'} \hat{p}^{\dagger}_{s_{z}} \hat{p}_{s_{z}'}
 T^{1 s_{z} I}_{1 s_{z}' L_{1} L_{2}}(\Omega_{q})
&=& \sum_{s_{z} s_{z}'} \hat{n}^{\dagger}_{s_{z}} \hat{n}_{s_{z}'}
 T^{1 s_{z} I}_{1 s_{z}' L_{1} L_{2}}(\Omega_{q}) \nonumber  \\
% &=& \sum_{l } ( 1 0 1 0 | l 0 ) ( L_{1} 0 L_{2} 0 | l 0 ) W( L_{1} 1 L_{2} 1
%%; I l )
%     \frac{2I+1}{4 \pi} \sqrt{(2L_{1}+1)(2L_{2}+1)} \nonumber \\
% & & \times  (-)^{I} P_{l}(0) \label{4V}
    &=& \frac{2I+1}{4 \pi} (-)^{I+1} \left\{ \right. \frac{1}{3} (-)^{L_{1}}
            \delta_{L_{1},L_{2}} \label{4V} \\
    & &  + \frac{1}{\sqrt{6}} (L_{1} 0 L_{2} 0 | 2 0) W(L_{1} 1 L_{2} 1; I 2)
            \sqrt{(2L_{1}+1)(2L_{2}+1)} \left. \right\}. \nonumber
\end{eqnarray}
The derivation of
Eq.(\ref{4V}) is in the appendix.
Thus the two spin-transverse response functions (\ref{4C}) and (\ref{4D})
are identical. \par
Now we write down the spin observables (\ref{1N})
in terms of
the scattering coefficients $ A \sim F $ and the spin response functions.
When we calculate $I_{0}$ and $D_{i}$, we have only those response functions in
Eq.(\ref{4A}) $\sim$ Eq.({\ref{4D}) and the interference term does not appear .
{}From Eqs.(\ref{2J}) and (\ref{4A}) $\sim$ (\ref{4D})
the unpolarized double differential cross section becomes
\begin{eqnarray}
 I_{0} &=& 48 K \zeta^{2} \left\{ (|A_{1}|^2 + |C_{1}|^2) R_{S}
            + (|B_{1}|^2 + |C_{1}|^2 + |F_{1}|^2) R_{T}
            + |E_{1}|^2 R_{L} \right\}.  \label{4W}
\end{eqnarray}
As for the polarization observables $D_{i}$ defined in Eqs.(\ref{1S}),
\begin{eqnarray}
 I_{0} D_{0} &=& 48 K \zeta^{2} \left( |A_{1}|^2  R_{S}
            + |C_{1}|^2 R_{T} \right) , \nonumber \\
 I_{0} D_{n} &=& 48 K \zeta^{2} \left( |C_{1}|^2 R_{S}
            + |B_{1}|^2 R_{T}  \right) , \nonumber \\
 I_{0} D_{p} &=& 48 K \zeta^{2} |F_{1}|^2 R_{T} , \nonumber \\
 I_{0} D_{q} &=& 48 K \zeta^{2} |E_{1}|^2 R_{L} . \label{4Y}
\end{eqnarray}
\par
\setcounter{equation}{0}
\section{Results and Discussions}
For the deuteron we used the wave function for the Reid soft core
potential \cite{reid}.
We solved the Schr\"odinger equation for $\psi^{JS}_{l',l}(r)$ in Eq.(\ref{3F})
with the
same
potential which is supplied for $J \leq 2$. For $J \geq 3$ we neglected the
nuclear
interaction and set
$\psi^{JS}_{l',l}(r) = F_{l}(kr) \delta_{l',l}$, where $F_{l}(x)$ is a
Coulomb wave function.
We used two different {\it NN} scattering amplitudes. One is provided by Bugg
and
Wilkin \cite{buggwilkin2}, and the other is calculated in the SAID system from
the phase shift SM89 of Arndt et al.\cite{arndt}
\par
In Fig.1 we show the results of the response functions
(a)$R_{S}$, (b)$R_{L}$ and (c)$R_{T}$. The solid lines
show the results with full correlations of the initial and the final target
state. The
dashed lines indicate the results with the final state correlations. The
results when the
deuteron consists of only $S$-wave are presented with the dotted lines.
The dash-dotted lines are the contributions of the final $S$-wave to
$R_{L}$ and $R_{T}$.
All three response functions have large
peaks at $\omega \approx 60$ MeV and their general features are almost the
same.
Comparing the solid line with the dashed line we see the effects of the final
$2p$
correlations. They commonly reduce the quasi-elastic peak and slightly shift
$R_{T}$ toward lower $\omega$. As is expected, the $S$-wave interaction of $2p$
forms a sharp peak
near the threshold in $R_{L}$ and $R_{T}$. For $R_{T}$ it is much
smaller than for $R_{L}$. This difference is due to the $D$-state of the
deuteron.
As we see in Fig.1(b) and (c), the threshold peak is almost the same in $R_{L}$
and
$R_{T}$ unless this $D$-state exists. From this we see the
interference term with
the initial $D$-state and final $S$-state makes opposite contributions to
$R_{L}$
and $R_{T}$. \par
Figure 2 shows the ratio $R_{L}/R_{T}$. Except for the dash-dotted line the
meaning of each line is the same as in Fig.1. In the dash-dotted line we
replaced
the final $P$-waves by those without correlations. The behavior of the solid
line
is essentially the same as the one obtained in Ref.\ \cite{pandhari}.
Comparing the solid and the
dashed lines we find that the ratio becomes smaller at the lower $\omega$
side and becomes close to
unity
at $\omega \approx 50$ MeV when we include full final state
interactions.  At higher $\omega$ side it increases
gradually up to 1.4, though the ratio remains
stable at 1.1 in the case of the uncorrelated final state. The interference
term with
the initial $D$-state and final $P$-state
contributes differently to $R_{L}$ and $R_{T}$, and the $P$-wave correlation
amplifies this feature. The initial $D$-state plays an
essential role in raising the ratio near the threshold.
\par
We display the results of the unpolarized double differential cross section
in the laboratory frame $I_{lab}$  in Fig.3 and
the polarization quantities $D_{0}$, $D_{n}$, $D_{p}$,
$D_{q}$ in Fig.4.
The solid lines and the dashed lines represent those obtained with and without
the
final state correlations respectively. They are calculated by use of the {\it
NN}
amplitude of Bugg and Wilkin \cite{buggwilkin2}. The dotted lines show the
results with the final state correlations, but calculated by the {\it NN}
amplitude
of Arndt et al.\cite{arndt}.
For $I_{lab}$,
the peak position
and its width is almost
the same as the
experimental result, but the theoretical magnitude is more than 10\% larger.
Comparing
the result of Bugg-Wilkin's amplitude with that of the Arndt's amplitude
we find the uncertainty caused by
the ambiguity of the {\it NN} amplitude. It is about 4 \% in this case.
So this is not the only reason for the deviation from the experimental result.
A small peak due to the $S$-wave
correlation of the final $2p$ appears near the threshold in solid and dotted
lines.
\par
The polarization quantities $D_{0}$, $D_{n}$, $D_{p}$, and $D_{q}$ are in
rather good
agreements with the experimental data in the range of errors. The calculation
by
the uncorrelated final state does not differ so much from the one with the
correlations. However the differences between the results by the two {\it NN}
amplitudes are distinct in $D_{n}$ and $D_{p}$.
The behavior of $D_{n}$, $D_{p}$ and $D_{q}$ can be understood with a
help of the ratio $R_{L}/R_{T}$. For example $D_{n}$ is written as
\begin{eqnarray}
D_{n} & = & \frac{|C_{1}|^2 R_{S}/R_{T} + |B_{1}|^2}
                       { (|A_{1}|^2 + |C_{1}|^2) R_{S}/R_{T}
                         + (|B_{1}|^2 + |C_{1}|^2 + |F_{1}|^2)
                         + |E_{1}|^2  R_{L}/R_{T} } \label{5B}
\end{eqnarray}
from Eqs.(\ref{4W}) and (\ref{4Y}).
The values $|A_{1}|^2$ and $|C_{1}|^2$ are considerably smaller than
$|B_{1}|^2$, $|E_{1}|^2$,
and $|F_{1}|^2$. The numerator is almost determined by $|B_{1}|^2$, and
$|E_{1}|^2  R_{L}/R_{T}$ is the main term which changes the denominator.
{}From the behavior of the ratio $R_{L}/R_{T}$ shown in Fig.2,
$D_{n}$ is expected to have a broad peak at
$\omega \approx 50$ MeV and a sharp decrease near the threshold.
We can derive the analogous equations for $D_{p}$ and $D_{q}$.
They show that $D_{p}$ is expected to
behave similarly as $D_{n}$ and $D_{q}$ to have
a broad dip at $\omega \approx 50$ MeV and a sharp rise near the threshold.
Calculated results have these features, though they are not apparently
seen in the experimental data. \par
On the whole our calculation agrees rather well with the experiment, but it
somewhat
overestimates $I_{lab}$. One reason of this discrepancy is the
ambiguity of the {\it NN} amplitude. We have seen that this causes the
uncertainty of
several percent in $I_{lab}$. There are also clear differences in $D_{n}$
and $D_{p}$ between the results by the two amplitudes.
Another reason is the applicability of the
optimal factorization. In Table II we present the values of the relevant
quantities.
The value of $\eta$ increases rapidly as $\omega$. On the contrary the kinetic
energy
of the projectile in the {\it NN} laboratory frame $T^{NN}_{lab}$ decreases
correspondingly.
The {\it NN} amplitude varies rather much as a function of this $T^{NN}_{lab}$.
We consider that the optimal factorization-like treatment is needed,
but there may be other ways of estimating the internal motion.
A naive treatment of the reaction is to assume that the
target nucleon is at rest in the target, which corresponds to the case of
$\eta = 1/4$ as is
seen from Eq.(\ref{1H}).
At $\omega=60$ MeV, $\eta$ is 0.265.
So this picture is partially realized
near the peak. We may attribute the overestimation of $I_{lab}$ largely to
neglecting the absorption of the flux, such as a shadow effect. Virtual
$\Delta$
excitation may also be a reason. \par
In summary the ratio $R_{L}/R_{T}$ is close to unity near the peak. Carey
et al.\cite{carey} used the quantities of the deuteron for reference by taking
average of all $\omega$. Since most of the data are taken near the peak, their
prescription of $R_{L}/R_{T}=1$ is permissible, or it should be slightly larger
if we want to fix it for convenience. We can also conclude that
PWIA is a reasonable approximation for the reaction $d(\vec{p},\vec{n})2p$.
It gives rather good results for $D_{i}$.
It also well reproduces the shape of $I_{lab}$. However it overestimates the
magnitude, which cannot be explained only by the uncertainty of the {\it NN}
amplitude. \\ \par
The numerical calculation in this work was performed on the TKYVAX node in
the Meson Science Laboratory, University of Tokyo. This work was supported by
Grants-in-Aid for Scientific Research of the Ministry of Education
(No.02640215, 05640328).
\par
\setcounter{equation}{0}
\renewcommand{\theequation}{A.\arabic{equation}}
\renewcommand{\thesection}{Appendix}
\section{}
{}From Eq.(\ref{4Q}) Eq.(\ref{4V}) is calculated as follows:
\begin{eqnarray}
 & & \sum_{s_{z} s_{z}'} \hat{n}^{\dagger}_{s_{z}} \hat{n}_{s_{z}'}
     T^{1 s_{z} I}_{1 s_{z}' L L'}(\Omega_{q}) \nonumber \\
 &=& \frac{4 \pi}{3} \sum_{s_{z} s_{z}'} \sum_{M_{L} M_{L'}} \sum_{M_{I}}
     ( L M_{L} 1 s_{z} | I M_{I} ) ( L' M_{L'} 1 s_{z}' | I M_{I} )
     \nonumber \\
 & & \times (-)^{s_{z}} Y_{1 \ -s_{z}}(\Omega_{n}) Y_{1 s_{z}'}(\Omega_{n})
     (-)^{M_{L}} Y_{L \ -M_{L}}(\Omega_{q}) Y_{L' M_{L'}}(\Omega_{q})
     \nonumber \\
 &=& \sum_{s_{z} s_{z}'} \sum_{M_{L} M_{L'}} \sum_{M_{I}} \sum_{l_{1} l_{2}}
     (-)^{ -s_{z} + s_{z}' + M_{L} - l_{1} + I - L' }
     \sqrt{\frac{4 \pi}{2l_{1}+1}} \sqrt{\frac{4 \pi}{2l_{2}+1}}
     \sqrt{\frac{2L+1}{4 \pi}} \sqrt{\frac{2L'+1}{4 \pi}} \nonumber \\
 & & \times Y_{l_{1} \ -s_{z}+s_{z}'}(\Omega_{n})
     Y_{l_{2} \ -M_{L}+M_{L'}}(\Omega_{q})
     ( 1 0 1 0 | l_{1} 0 ) ( L 0 L' 0 | l_{2} 0 ) \nonumber \\
 & & \times ( L \ -M_{L}\ L' M_{L'} | l_{2} \ -M_{L}+M_{L'} )
            ( L M_{L} 1 s_{z} | I M_{I} ) ( I M_{I} 1 \ -s_{z}' | L' M_{L'} )
     \nonumber \\
 & & \times ( 1 s_{z} 1 \ -s_{z}' | l_{1} \ s_{z}-s_{z}' ) \nonumber \\
 &=& \sum_{l m} ( 1 0 1 0 | l 0 ) ( L 0 L' 0 | l 0 ) W( L 1 L' 1 ; I l )
     \frac{2I+1}{2l+1} \sqrt{(2L+1)(2L'+1)} \nonumber \\
 & & \times  (-)^{I+m} Y_{l -m}(\Omega_{n})
     Y_{l m}(\Omega_{q}) \label{6A}
\end{eqnarray}
where we use the relation
\begin{eqnarray}
   \lefteqn{ \sqrt{(2e+1)(2f+1)} ( a \alpha f \ \gamma-\alpha | c \gamma )
    W( a b c d;e f) } \nonumber \\
    &=& \sum_{\beta} ( a \alpha b \beta | e \ \alpha+\beta )
                     ( e \ \alpha+\beta \ d \ \gamma-\alpha-\beta | c \gamma )
                     ( b \beta d \ \gamma-\alpha-\beta | f \ \gamma-\alpha ).
                     \label{6B}
\end{eqnarray}
In Eq.(\ref{6A}) we can sum over $m$ by using the identity
\begin{equation}
 \sum_{m} Y^{\ast}_{l m}(\Omega_{n}) Y_{l m}(\Omega_{q}) = \frac{2l+1}{4 \pi}
 P_{l}(\hat{\bold{n}} \cdot \hat{\bold{q}}) = \frac{2l+1}{4 \pi} P_{l}(0).
 \label{6C}
\end{equation}
Due to the presence of $(1 01 0 | l 0)$, we have only to take the sum over
$l=0,2$.
After inserting the explicit values of $P_{l}(0)$ we obtain Eq.(\ref{4V}).
\newpage
\pagestyle{empty}
%
%Tables
%
\noindent
\hspace*{85pt}
Table I. \ $\psi$ which is the angle between $\bold{q}$ and $\bold{q}_{c}$ and
\\
\hspace*{95pt}
the spin rotation angles $\chi$, $\chi'$, $\rho$ and $\rho'$.
\[
\begin{array}{c | rrrrrl} \hline
 \omega &    40 &    60  &    80  &   100 &   120 & \mbox{MeV} \\ \hline
 \psi   & 0.072 & 0.222  & 0.363  & 0.492 & 0.605 & \mbox{deg} \\
 \chi   & 0.936 & -0.016 & -0.911 & -1.726 & -2.440 & \\
 \chi'  & 1.826 & 2.279  & 3.579 & 4.353 & 5.028 & \\
 \rho   & 0.967 & -0.017 & -0.933 & -1.756 & -2.464 & \\
 \rho'  & 1.891 & 2.824  & 3.697 & 4.481 & 5.152 & \\ \hline
\end{array}
\]
\\ \\
\hspace*{95pt}
Table II. \ The values related to the optimal factorization: \\
\hspace*{105pt}
$\eta$, the corresponding incident energy in the {\it NN} labora-\\
\hspace*{105pt}
tory frame $T^{NN}_{lab}$ and the scattering angle $\theta^{NN}_{lab}$.
\[
\begin{array}{c | rrrrl} \hline
 \omega_{lab} &    40  &    60  &    90   &   120 & \mbox{MeV} \\ \hline
 \eta   & 0.090 & 0.265  & 0.519  & 0.751 &  \\
 T^{NN}_{lab} & 513 & 493 & 456 & 413  & \mbox{MeV}\\
 \theta^{NN}_{lab}  & 17.7 &  18.0 & 18.9 &  20.2 & \mbox{deg}\\ \hline
\end{array}
\]
\newpage
%
%Figure caption
%
\noindent
Fig.1. Response functions (a) $R_{S}$, (b) $R_{L}$ and
(c) $R_{T}$. The solid line includes full
correlations.
The dashed line has no correlations in the final state.
The dotted line is the result of purely $S$-wave deuteron.
The dash-dotted line indicates the contribution of the final $S$-wave. \\ \\
Fig.2. Ratio $R_{L}/R_{T}$.
Lines are the same as in Fig.1 except for the dash-dotted line.
The dash-dotted line has no correlations in the final P-waves. \\ \\
Fig.3. Unpolarized double differential cross section in the
laboratory frame $I_{lab}$.
The solid line includes full correlations and
the amplitude of Bugg and Wilkin is used.
The dashed line does not include final state interactions and the amplitude
of Bugg and Wilkin is used.
The dotted line includes full correlations and
the amplitude of Arndt et al. is used. The experimental data in
Ref.\ \cite{chen} are
represented by the dots. \\ \\
Fig.4. Polarization quantities $D_{0}$, $D_{n}$, $D_{p}$ and
$D_{q}$. Lines are the same as in Fig.3.
The experimental data in Ref.\ \cite{chen} are represented by the dots
with the error bars.
\end{document}